\documentclass[intlimits,twoside,a4paper]{article}

\usepackage{amsmath,amssymb}
\usepackage{graphicx}
\usepackage{wrapfig}

\usepackage[T2A]{fontenc}
\usepackage[cp1251]{inputenc}

\usepackage[eqsecnum]{cmpj2}

\issue{2012}{15}{4}{43603}

\doinumber{10.5488/CMP.15.43603}

\newcommand {\rv}{{\bf r}}
\newcommand{\Trcl}{\mathrm{Tr}}
\newcommand{\Fcal}{\mathcal{F}}
\newcommand{\Tr}{\mathrm{Tr}}

\newcommand {\xv}{{\bf x}}

\newcommand{\myfigwidth}{0.55\columnwidth}

\title[Recent developments in classical density functional theory]
      {Recent developments in classical density functional theory:
        Internal energy functional and diagrammatic structure of
        fundamental measure theory}

\author[M.~Schmidt \textsl{et al.}]{M.~Schmidt\refaddr{bt,bs}, M.~Burgis\refaddr{bt},
  W.S.B.~Dwandaru\refaddr{bs,indonesia}, G.~Leithall\refaddr{bs}, P.~Hopkins\refaddr{bs}}

\authorcopyright{M.~Schmidt, M.~Burgis, W.S.B.~Dwandaru, G.~Leithall, P.~Hopkins, 2012}

\addresses{
\addr{bt} Theoretische Physik II, Physikalisches Institut,
  Universit{\"a}t Bayreuth, D--95440 Bayreuth, Germany
\addr{bs} H.H.\ Wills Physics Laboratory, University of Bristol,
   Tyndall Avenue, Bristol BS8 1TL, UK
\addr{indonesia} Jurusan Fisika, Universitas Negeri Yogyakarta,
  Bulaksumur, Yogyakarta, Indonesia
}

\date{Received August 15, 2012}
\begin{document}
\maketitle

\begin{abstract}
  An overview of several recent developments in density functional
  theory for classical inhomogeneous liquids is given. We show how
  Levy's constrained search method can be used to derive the
  variational principle that underlies density functional theory. An
  advantage of the method is that the Helmholtz free energy as a
  functional of a trial one-body density is given as an explicit
  expression, without reference to an external potential as is the
  case in the standard Mermin-Evans proof by reductio ad absurdum. We
  show how to generalize the approach in order to express the internal
  energy as a functional of the one-body density distribution and of
  the local entropy distribution. Here the local chemical potential
  and the bulk temperature play the role of Lagrange multipliers in
  the Euler-Lagrange equations for minimiziation of the functional. As
  an explicit approximation for the free-energy functional for hard
  sphere mixtures, the diagrammatic structure of Rosenfeld's
  fundamental measure density functional is laid out. Recent
  extensions, based on the Kierlik-Rosinberg scalar weight functions,
  to binary and ternary non-additive hard sphere mixtures are
  described.
  \pacs 61.25.-f, 61.20.Gy, 64.70.Ja
\end{abstract}

\section{Introduction}

The theoretical study of inhomogeneous classical liquids received a
boost through the development of classical density functional theory
(DFT).  Evans' 1979 article~\cite{evans79} constitutes a central
reference to DFT. The proof of the variational principle that
underlies the theory, including the existence and the uniqueness of
the free energy functional, can be viewed as the classical analogue of
Mermin's earlier (1965) work on quantum systems at finite temperatures
\cite{mermin65}. This forms a generalization of the Hohenberg-Kohn
theorem for ground state properties of quantum systems. Historic
milestones that predate these developments, and that can be
re-formulated in classical DFT language, are Onsager's 1949 treatment
of the isotropic-nematic liquid crystal phase transition in systems of
long and thin hard rods~\cite{onsager49}, and van der Waals' 1893
theory of the microscopic structure of the liquid-gas interface
\cite{vanderwaals93}.

Both the Hohenberg-Kohn proof and the Mermin-Evans proof start from a
variational principle for the respective many-body function. In the
quantum case, this is the Rayleigh-Ritz inequality for the many-body
groundstate wave function. In the classical case, the theory rests on
the Gibbs inequality for the equilibrium many-body phase space
distribution. The corresponding functionals are the ground state
energy in the quantum case and the thermodynamic grand potential in
the classical case. Both functionals depend (trivially) on the
position-dependent external one-body potential. Via an intricate
sequence of arguments~\cite{mermin65,evans79}, which has become
textbook knowledge~\cite{hansen06}, the dependence on the external
potential is played back to the more useful dependence on the (in
general) position-dependent one-body density distribution.

In 1979 Levy showed that quantum DFT can be obtained in a more compact
and straightforward way via a method that he called the constrained
search~\cite{levy79} (see also~\cite{percus78}). Here, the search
for the minimum is performed in the space of all trial many-body wave
functions. The constraint that is fixed during this search is that all
trial wave functions considered need to generate the {\em given}
one-body density distribution. Levy's derivation is both rigorous and
elegant and constitutes a standard reference for electronic structure
DFT. Among the impressive number of citations of his 1979 article,
there are only very few papers that draw connections to classical DFT,
~\cite{weeks03} being an example, although the method permits
rather straightforward application to the classical case
\cite{dwandaru11levy}.  See also Percus' general concept of
``overcomplete'' density functionals~\cite{percus98}.


The constrained search method offers two significant benefits over the
Mermin-Evans proof. One is simplicity, avoiding the reductio ad
absurdum chain of arguments. The other is that it yields an explicit
definition of the Helmholtz free energy density as a functional of the
(trial) one-body density distribution. This formula, as reproduced in
(\ref{EQFviaLevy}) below, is explicitly independent of the external
potential. The generalization that facilitates this development is the
concept of minimization in the space of all phase space distributions
under the constraint of a given one-body density (\ref{EQfTorho}).
The issue of representability of the one-body density has been addressed
in~\cite{chayes84}.  It turns out that the simplicity of Levy's
method allows one to construct more general DFTs. An example is the
variational framework developed in~\cite{schmidt11edft}, which rests
on the {\em internal} energy functional (rather than the Helmholtz
free-energy functional), which depends on the one-body density and on
a local (position-dependent) entropy distribution. A dynamical version
of this theory, based on linear irreversible thermodynamics and
phenomenological reasoning, is proposed in
\cite{schmidt11edft}. Several common approximations for free energy
functionals were transformed to internal energy functionals. Having
reliable approximations for the functional is a prerequisite for
applying DFT to realistic (three-dimensional) model systems. The task
of constructing usable functionals is different from the conceptual
work outlined so far. In particular, a calculation of the constrained
search expressions for the (free-energy or internal-energy)
functional would amount to an exact solution of the many-body
problem. Hence, the importance of these expressions is rather of
conceptual nature.

In 1989 Rosenfeld wrote a remarkable letter in which he proposed an
approximate free energy functional for additive hard sphere mixtures
\cite{rosenfeld89}. His theory unified several earlier liquid state
theories, such as the Percus-Yevick integral equation theory for the
bulk structure~\cite{hansen06}, the scaled-particle theory for
thermodynamics, and Rosenfeld's own concepts, such as the scaled-field
particle theory of~\cite{rosenfeld89}, and encapsulated these
into what he called fundamental measure theory (FMT). Kierlik and
Rosinberg in 1990~\cite{kierlik90} re-wrote the same functional
\cite{Phan93} in an alternative way, using only four (not six, as
Rosenfeld) weight functions to build weighted densities via
convolution with the bare one-body density of each hard sphere
species. The two strands of FMT were pursued both with significant
rigour and effort, see the recent reviews
\cite{tarazona08review,roth10review,lutsko10review}. Rosenfeld's more
geometric approach was extended to further weight functions by
Tarazona~\cite{tarazona00} in his treatment of freezing. A critical
discussion of the properties of the relevant convolution kernels can
be found in~\cite{cuesta02}. Very recently, Korden
\cite{korden2012virial,korden2012loops} demonstrated the relationship
of the FMT with the exact virial expansion.

It is vexing that Rosenfeld himself, who was certainly very versed in
the diagrammatic techniques of liquid state theory (see
e.g.~\cite{rosenfeld88}),  apparently neither analysed nor formulated
his very own FMT within such a framework.  The non-local structure of
FMT, its coupling of space integrals via convolution, and the
plentiful appearances of the one-body density distribution(s) seem to
constitute an ideal playground for formulation in diagrammatic
language. A comprehensive understanding of the diagrammatic nature of
FMT could not  help just to ascertain and clarify the nature of the
approximations that are involved~\cite{korden2012virial}, but also,
and more importantly from a pragmatic point of view, could enable one to
construct new functionals for further model systems. Taking the FMT
weight functions as bonds in a diagrammatic formulation, one
immediately faces the combinatorial problems associated with their
number, which in the Rosenfeld-Tarazona formulation are at least seven
per hard sphere species (four scalar, two vector, one tensor), which
makes the book-keeping task a seemingly daunting one.

It was shown~\cite{leithall11hys,leithall12} that one can formulate
the Kierlik-Rosinberg form of FMT in a diagrammatic way. The concept
was applied to one-dimensional hard rods, where it gives Percus' exact
result~\cite{percus76,vanderlick89}, as well as to five-dimensional
hard hypersphere mixtures, where it gives a functional that
outperforms Percus-Yevick theory, and improves on previous FMT
attempts~\cite{finken02dhs}. Two crucial properties of the
diagrammatic formulation can be identified. One is the relative
simplicity of the diagrams that describe the coupling of the various
space integrals in the density functional.  The topology of the
diagrams is of star-like or tree-like shape, hence providing a  significant
reduction as compared to the complexity of the exact virial series. The
number of arms is equal to the power in density and the bonds are
weight functions rather than Mayer functions, like they are in the exact
virial expansion.  The book-keeping problem of having to deal with a
large number of different weight functions is addressed and rendered
almost trivial by exploiting the tensorial structure that underlies
the Kierlik-Rosinberg form of FMT~\cite{schmidt11nagl}, where the
geometric index of the weight functions is a proper tensorial index,
and corresponding (isometric and metamorphic) transformations can be
formulated~\cite{schmidt11nagl}. Hence, the fully scalar
Kierlik-Rosinberg formulation turned out to be indeed simpler to
handle, and easier to generalize for full control of the degree of
non-locality in the functional. These developments facilitated the
generalization of FMT for binary non-additive hard spheres
\cite{schmidt04nahs} to ternary mixtures~\cite{schmidt11tnas}.

In the present contribution, we describe the basic ideas underlying
the above developments, without the full detail that is given in the
respective original papers, but with further illustrative examples in
order to provide an introduction to the subject. The paper is
organized as follows. In section~\ref{SEClevy}, Levy's method is
sketched, and both the free-energy and the internal-energy functionals
are defined. A basic introduction to the diagrammatic formulation of
FMT is given in section~\ref{SECdiagrammatic}, including a brief
overview of applications to non-additive hard sphere fluids in bulk
and at interfaces. Section~\ref{SECconclusions} gives concluding
remarks.

\section{Levy's constrained search in classical DFT}
\label{SEClevy}
\subsection{Variational principle for the grand potential functional}

Consider a classical many-body system with position coordinates
$\rv_1,\rv_2,\ldots,\rv_N$ and momenta \linebreak ${\bf p}_1, {\bf p}_2, \ldots,
{\bf p}_N$, where $N$ is the number of particles. Denote the classical
trace in the grand ensemble by
\begin{equation}
  \Trcl = \sum^\infty_{N=0}
  \frac{1}{h^{3N}N!}\int \rd\rv_1\ldots \rd\rv_N
  \int \rd{\bf p}_1\ldots \rd{\bf p}_N\,,
\end{equation}
where $h$ is Planck's constant. For a given total interatomic potential
$U(\rv_1,\ldots,\rv_N)$ between the particles, one {\em defines} the
intrinsic Helmholtz free energy functional by the explicit
expression~\cite{dwandaru11levy}
\begin{equation}
  \Fcal[\rho] =
  \min_{f\rightarrow\rho}
  \left[\Trcl f\left(\sum^N_{i=1}\frac{p^{2}_{i}}{2m}+U+
    k_{\mathrm{B}}T\ln f\right)\right],
  \label{EQFviaLevy}
\end{equation}
where $T$ is temperature, $k_{\mathrm{B}}$ is the Boltzmann constant, $m$ is the
particle mass, $p_i=|{\bf p}_i|$, and the minimization searches all
many-body (phase space) probability distributions
$f(\rv_1,\ldots,\rv_N,{\bf p}_1,\ldots,{\bf p}_N; N)$ that are
normalized according to
\begin{equation}
  \Trcl f = 1,
\end{equation}
and that yield the fixed trial one-body density $\rho(\rv)$ via
\begin{equation}
  \rho(\rv) = \Trcl f \sum_{i=1}^N\delta(\rv-\rv_i).
  \label{EQfTorho}
\end{equation}
The relationship (\ref{EQfTorho}) is indicated as $f\rightarrow \rho$
in the notation of (\ref{EQFviaLevy}). It can be shown
\cite{dwandaru11levy} that the intrinsic free energy functional
(\ref{EQFviaLevy}) satisfies the Euler-Lagrange equation
\begin{equation}
  \left.\frac{\delta  \Fcal[\rho]}{\delta \rho(\rv)}
  \right|_{\rho_0} = \mu - v(\rv),
  \label{EQEulerLagrange}
\end{equation}
where $\mu$ is the chemical potential, $v(\rv)$ is an external
one-body potential acting on the system, so that the total external
potential for a given microstate is $\sum_{i=1}^N v(\rv_i)$, and
$\rho_0(\rv)$ is the equilibrium one-body density,
\begin{equation}
  \rho_0(\rv) = \Trcl f_0 \sum_{i=1}^N\delta(\rv-\rv_i),
  \label{EQrho0}
\end{equation}
where the equilibrium many-body distribution $f_0$ is given by the
Boltzmann distribution,
\begin{align}
    f_0 = \Xi^{-1} \exp\left(-\frac{H_N-\mu N}{k_{\mathrm{B}}T}\right).
  \label{EQf0}
\end{align}
Here, the normalization constant is the grand partition sum
\begin{align}
  \Xi = \Tr \exp\left(-\frac{H_N-\mu N}{k_{\mathrm{B}}T}\right)
\end{align}
and the Hamiltonian is
\begin{align}
   H_N = \sum_{i=1}^N \frac{p_i^2}{2m} + U(\rv_1,\ldots,\rv_N) +
   \sum_{i=1}^N v(\rv_i).
   \label{eq:2.9}
\end{align}
A concise version of the proof of (\ref{EQEulerLagrange}) is laid out
in the following.

\subsection{Sketch of the constrained search proof a la Levy}
\label{SECtwoStageMinimization}
It can easily be shown~\cite{evans79} that the equilibrium grand
potential $\Omega_0$ for Hamiltonian $H_N$ is obtained from minimizing
Mermin's form of the grand potential functional,
\begin{equation}
  \Omega_0 = \min_f \Trcl f \left(H_N-\mu N + k_{\mathrm{B}}T \ln f\right),
\end{equation}
where $f$ is a trial many-body distribution.  Decompose the right hand
side into a double minimization
\begin{equation}
  \Omega_0 = \min_\rho \min_{f\rightarrow\rho}
  \Trcl f \left(H_N-\mu N + k_{\mathrm{B}}T \ln f \right),
    \label{EQomega0asDoubleMinimization}
\end{equation}
where the inner minimization is a search under the constraint that the
$f$ generates $\rho$ via (\ref{EQfTorho}).  For suitable Hamiltonians as (\ref{eq:2.9})
with external potential $v(\rv)$, this can be written as
\begin{eqnarray}
  \Omega_0 = \min_\rho \min_{f\rightarrow\rho}
  \Trcl f \left(\sum_{i=1}^N\frac{p_i^2}{2m} + U
  + \sum_{i=1}^N v(\rv_i)
  -\mu N + k_{\mathrm{B}}T \ln f \right).
  \label{EQomega0explicitHamiltonian}
\end{eqnarray}
In the expression above
\begin{equation}
  \Trcl f \left[\sum_{i=1}^N v(\rv_i) - \mu N\right] =
  \int \rd\rv\left[v(\rv)-\mu\right]\rho(\rv),
\end{equation}
because $f\rightarrow\rho$. So we may re-write
(\ref{EQomega0explicitHamiltonian}) as
\begin{eqnarray}
  \Omega_0 = \min_\rho \left\{ \int \rd\rv \left[v(\rv)-\mu\right]\rho(\rv)
  +\min_{f\rightarrow\rho}
    \Trcl f\left(\sum_{i=1}^N\frac{p_i^2}{2m}+U+k_{\mathrm{B}}T\ln f\right)
  \right\}\nonumber
\end{eqnarray}
or
\begin{equation}
 \Omega_0 = \min_\rho \left\{
  \int \rd\rv \left[v(\rv)-\mu\right]\rho(\rv) + \Fcal[\rho]
 \right\},
 \label{EQomega0asMinimization}
\end{equation}
where $\Fcal[\rho]$ is given by (\ref{EQFviaLevy}) and hence in
equilibrium the Euler-Lagrange equation (\ref{EQEulerLagrange})
follows.
Finally, note that from (\ref{EQomega0asDoubleMinimization}) the grand
potential density functional can be defined as
\begin{align}
  \Omega[\rho] =
  \min_{f\rightarrow\rho} \left[
  \Trcl f \left(H_N-\mu N + k_{\mathrm{B}}T \ln f \right) \right].
\end{align}

\subsection{DFT based on the internal-energy functional}
Using two constraints, rather than one, one can define the
internal-energy functional as
\begin{align}
  E[\rho,s] = \min_{f\to\rho,s} \left\{f \left[\sum_{i=1}^N\frac{p_i^2}{2m}
  + U(\rv_1,\ldots,\rv_N) \right]\right\},
  \label{EQinternalEnergyFunctional}
\end{align}
where the constraint for the density is (\ref{EQfTorho}) and that for
the local entropy distribution $s(\rv)$ is
\begin{align}
  s(\rv) = -k_{\mathrm{B}}\Tr \frac{1}{N}
  \sum_{i=1}^N \delta(\rv-\rv_i)
  f\ln f. \label{EQfTOs}
\end{align}
The equilibrium value for the local entropy is
\begin{align}
  s_0(\rv) = -k_{\mathrm{B}}\Tr \frac{1}{N}
  \sum_{i=1}^N \delta(\rv-\rv_i)
  f_0\ln f_0\,, \label{EQs0}
\end{align}
with the equilibrium many-body distribution $f_0$ given in
(\ref{EQf0}), and the total entropy obtained as the spatial integral
$S_0=\int \rd\rv s_0(\rv)$.  The Euler-Lagrange equations have the form
\begin{align}
  \left.\frac{\delta E[\rho,s]}{\delta \rho(\rv)}
  \right|_{\rho_0,s_0} &= \mu - v(\rv),
  \label{EQELrho}\\
  \left.\frac{\delta E[\rho,s]}{\delta s(\rv)}
  \right|_{\rho_0,s_0} &= T.
  \label{EQELs}
\end{align}
Explicit forms of various internal-energy functionals can be found in
\cite{schmidt11edft}, as can be a dynamic prescription, similar in
spirit to dynamical DFT~\cite{archer04ddft}, for the joint time
evolution of $s(\rv,t)$, $\rho(\rv,t)$, where $t$ is time. The proof
of (\ref{EQinternalEnergyFunctional})--(\ref{EQELs}) is a
straightforward application of Levy's method, as laid out in section
\ref{SECtwoStageMinimization}; for the explicit form see
\cite{schmidt11edft}.

\section{Fundamental measure theory for hard sphere mixtures}
\label{SECdiagrammatic}
\subsection{Diagrammatic formulation for additive and non-additive mixtures}
We turn to hard sphere mixtures and use the common splitting of the
intrinsic Helmholtz free energy into ideal gas and excess (over ideal
gas) contributions
\begin{align}
  F[\{\rho_i\}] = k_{\mathrm{B}}T \sum_i\int \rd \rv
  \rho_i(\rv)\left\{\ln\left[\rho_i\left(\rv\right)\Lambda_i^3\right]-1\right\} + F_{\rm exc}[\{\rho_i\}],
\end{align}
where $\rho_i(\rv)$ is the one-body density distribution, $\Lambda_i$
is the thermal de Broglie wavelength of species $i$, and $F_{\rm
  exc}[\{\rho_i\}]$ is the Helmholtz excess free energy functional
that is due to interparticle interactions.  Its lowest order (in
density) virial expansion is
\begin{align}
  F_{\rm exc}[\{\rho_i\}] \rightarrow -
  \frac{k_{\mathrm{B}}T}{2}\sum_{i,j} \int \rd\rv \rd\rv'
  \rho_i(\rv)f_{ij}(\rv-\rv')\rho_j(\rv'),
  \label{EQfexc2ndOrder}
\end{align}
where the Mayer function is defined as
$f_{ij}(r)=\exp[u_{ij}(r)/(k_{\mathrm{B}}T)]-1$, with $u_{ij}(r)$ being the pair
interaction potential between species $i$ and $j$. The sum in
(\ref{EQfexc2ndOrder}) is over all species $i,j$.  For hard spheres
$f_{ij}(r<\sigma_{ij})=-1$ and zero otherwise; here $\sigma_{ij}$ is
the hard core interaction distance between species $i$ and $j$.

Kierlik and Rosinberg introduced a set of four ``weight functions''
$w_0(R_i,r), w_1(R_i,r), w_2(R_i,r)$, and $w_3(R_i,r)$, where
$R_i=\sigma_{ii}/2$ is the particle radius, and $r$ is radial distance
\cite{kierlik90}. Using the weight functions, the hard sphere Mayer
function can be written as a sum of convolution integrals,
\begin{align}
  -f_{ij}(r)  & =
  w_0(R_i) \ast w_3(R_j) +
  w_1(R_i) \ast w_2(R_j) +
  w_2(R_i) \ast w_1(R_j) +
  w_3(R_i) \ast w_0(R_j),
  \label{EQdeconvolutionKR}
\end{align}
where the asterisk denotes the convolution of two functions $h_1(r)$
and $h_2(r)$, which is defined as $(h_1\ast h_2)(|\rv-\rv'|)=\int \rd\xv
h_1(\rv-\xv) h_2(\rv'-\xv)$.  The spatial arguments of the weight
functions have been omitted in the notation of the right hand side of
(\ref{EQdeconvolutionKR}).  The direct space expressions of the weight
functions are $w_0(R,r)=-\delta''(R-r)/(8\pi)+\delta'(R-r)/(2\pi r)$,
$w_1(R,r)= \delta'(R-r)/(8\pi)$, $w_2(R,r)=\delta(R-r)$, and
$w_3(R,r)=\Theta(R-r)$, where $\delta(\cdot)$ is the Dirac
distribution, and $\Theta(\cdot)$ is the Heaviside step function.

The appearance of particular combinations of products in
(\ref{EQdeconvolutionKR}) can be based on dimensional analysis. In
order to see this, note that the weight functions $w_\nu$ are objects
with dimensions of $(\rm length)^{\nu-3}$, hence $w_0, w_1, w_2$, and
$w_3$ carry dimensions of $(\rm length)^{-3}, (\rm length)^{-2}, (\rm
length)^{-1}$, and $(\rm length)^0$, respectively. Each of the products
in (\ref{EQdeconvolutionKR}) has the dimension $({\rm length})^{-3}$,
which cancels the $({\rm length})^3$ which is due to the convolution
integral and hence yields a dimensionless Mayer function.

This property can be formalized~\cite{schmidt07nage} in order to
establish more mathematical structure. One can re-write
(\ref{EQdeconvolutionKR}) as
\begin{align}
  -f_{ij}(\rv-\rv') =
  \int \rd\xv
  \sum_{\nu,\tau=0}^3 M^{\nu\tau}
   w_\nu(R_i,\rv-\xv) w_\tau(R_j,\rv'-\xv),
   \label{EQdeconvolutionM}
\end{align}
where the coefficients $M^{\nu\tau}$ are chosen in such a way that
only the ``allowed'' combinations of weight functions contribute.
Hence, for most of the index combinations $\nu\tau$, the coefficients
vanish, $M^{\nu\tau}=0$.  Those that contribute in
(\ref{EQdeconvolutionKR}) possess the index combinations
$\nu\tau=03,12,21,31$, hence the non-vanishing coefficients are
$M^{03}=M^{12}=M^{21}=M^{30}=1$.  This procedure amounts to (only) a
formalization of (\ref{EQdeconvolutionM}). The crucial step is to view
the coefficients $M^{\nu\tau}$ as the elements of a metric $\sf M$
which operates in the space of weight functions ${\sf w}(R,\rv)$. Here,
a vector of weight functions $\sf w$ is defined by its components
$w_\nu(R,\rv)$, hence ${\sf w}=(w_0,w_1,w_2,w_3)$.  The length scale
$R$ acts as a parameter.  As an aside, one should not confuse the
vector $\sf w$ with Rosenfeld's vectorial weight functions ${\bf
  w}_{v1}(\rv)$ and ${\bf w}_{v2}(\rv)$~\cite{rosenfeld89}, which have
a very different origin, rooted in the geometry of the sphere in
three-dimensional space. The ${\sf w}(R,\rv)$, on the other hand, are
elements of a {\em four-dimensional} vector space; their index runs
from 0 to 3.

The metric {\sf M} can be represented as a matrix, which is akin to a
mirrored $4\times 4$ unit matrix in that its only non-vanishing
elements are unities on the {\em counter} diagonal,
\begin{align}
  {\sf M} \equiv
  \left(\begin{tabular}{cccc}
      0 & 0 & 0 & 1 \\
      0 & 0 & 1 & 0 \\
      0 & 1 & 0 & 0 \\
      1 & 0 & 0 & 0
  \end{tabular}\right).
  \label{EQmMatrix}
\end{align}

This formalization allows us to write (\ref{EQdeconvolutionM}) [and
  hence (\ref{EQdeconvolutionKR})] more concisely as follows:
\begin{align}
  f_{ij} = \int \rd\xv \;
  {\sf w}(R_i)\cdot {\sf M} \cdot {\sf w}^t(R_j),
\end{align}
where the superscript $t$ indicates matrix transposition, and the
spatial arguments have been omitted in the notation; these are the
same as in (\ref{EQdeconvolutionM}).

\begin{figure}[ht]
\centering
\includegraphics[width=\myfigwidth]{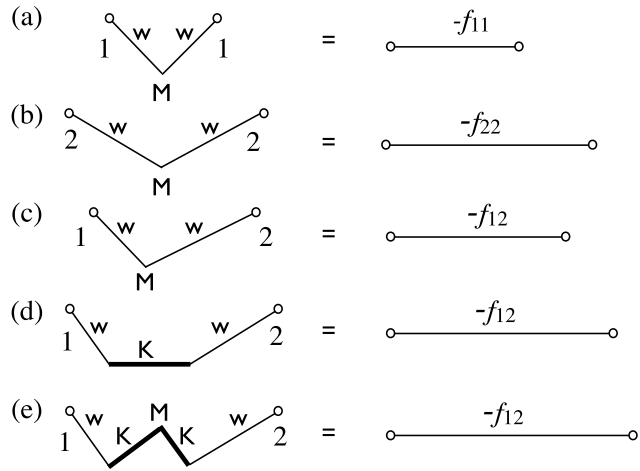}
\caption{Diagrammatic representation of the deconvolution of the Mayer
  bond $f_{ij}(r)$ between species $i,j=1,2$ for different types of
  hard sphere mixtures. The vector of species-specific weight
  functions $w_\nu(R_i,\rv)$, where $\nu=0,1,2,3$, is denoted by $\sf
  w$. Here, the species index $i=1,2$ is given at the root point (open
  circle).  Tensor contraction over pairs of the geometric (Greek)
  index with the metric $\sf M$ is implied, where $\sf M$ is a
  ``reversed unit matrix'', i.e., $M^{\nu\tau}=M_{\nu\tau}=1$ if
  $\nu+\tau=3$ and 0 otherwise (\ref{EQmMatrix}). Root points (open
  circles) are fixed spatial positions. Examples are shown for the KR
  deconvolution for species 11 (a), 22 (b), and 12 (c) in an additive
  mixture. For non-additive mixtures, the like-species expressions (a)
  and (b) still hold, but the cross-species expressions also possess
  the kernel matrix $\sf K$ (thick line), as defined in
  (\ref{EQkernelFromGenerator}) and (\ref{EQgenerator}) with
  index-raised elements $K^{\nu\tau}(d,r)$ as an additional type of
  bond. This appears as a single bond (d) in the binary
  non-additive functional, and as a double link in the ternary
  non-additive functional (e), here tensor-contracted by the metric $\sf
  M$. This allows one to control the range of $f_{12}(r)$ arbitrarily,
  i.e., independently of $R_1$ and $R_2$.}
\label{FIGone}
\end{figure}

In order to illustrate the framework, we display corresponding
diagrams for binary hard sphere mixtures in figure~\ref{FIGone}. The
deconvolution of the Mayer bond $f_{11}(r)$ between particles of
species 1 is displayed in figure~\ref{FIGone}~(a). The length of the
$\sf w$ bonds indicates the magnitude of the argument $R$, in this
case $R_1$. The kink is located at position $\xv$, which is the
integration variable in the convolution integral
(\ref{EQdeconvolutionM}). Hence, in this and in the following diagrams,
the position of a kink (or junction to be introduced below) is
integrated over. The open circles indicate fixed positions $\rv$ and
$\rv'$, i.e., the arguments in $f_{11}(|\rv-\rv'|)$. One commonly
refers to these as {\em root} points. Recall that multiplying each
root point by the one-body density distribution and integrating over
its position yields (up to a factor of $k_{\mathrm{B}}T/2$) the $ij=11$
contribution to the exact low density limit of the excess free energy
functional (\ref{EQfexc2ndOrder}). Figure~\ref{FIGone}~(b) gives the
corresponding diagram for species 2, here taken to be of a larger size
(hence $R_2>R_1$). The cross species diagram, $ij=12$, is shown in
figure~\ref{FIGone}~(c). From the diagrammatic representation it is
clear that the total length of the diagram, i.e., the range over which
$f_{12}(r)$ is non-zero, is fully determined by the accumulated length
of the two arms, $R_1$ and $R_2$. Hence, the cross species interaction
distance is $\sigma_{12}=R_1+R_2$, a case to which one refers to as an
{\em additive} hard sphere mixture.

Additivity is a fundamental property of the space of weight functions,
and is indeed more general than outlined so far. Consider a given
vector of weight functions ${\sf w}(R,\rv)$, where the length scale
$R$ is fixed. We aim at finding a linear operation ${\sf K}(d)$, where
$d$ is a length scale, that yields
\begin{align}
  {\sf w}^t(R') = {\sf K}(d)\; \dot{\ast} \; {\sf w}^t(R),
  \label{EQwTransport}
\end{align}
where $\dot\ast$ represents the combined operation of spatial
convolution ($\ast$) and dot product ($\cdot$) in the four-vector
space. Hence, in a less concise but more explicit way
(\ref{EQwTransport}) can be written componentwise as
\begin{align}
  w_\nu(R',\rv-\rv') = \int \rd\xv' \sum_{\nu=0}^3
  K_\nu^\tau(d,\rv-\xv) w_\tau(R,\rv'-\xv),
\end{align}
where the components of $\sf K$ are denoted by $K_\nu^\tau$ and the
two indices $\nu$ and $\tau$ run from 0 to 3. That such a $\sf K$
exists is a non-trivial matter, and forms the heart of the binary
non-additive hard sphere functional~\cite{schmidt04nahs}, see the
detailed investigation of the properties of $\sf K$ in
\cite{schmidt07nage}. The existence and properties are most easily
demonstrated in Fourier space representation, where the convolutions
become mere products. The Fourier transform of the ``kernel matrix''
$\sf K$ possesses a representation as the matrix exponential
\begin{align}
  \tilde {\sf K}(R,q) &= \exp
  \left(R{\sf G}\right),
  \label{EQkernelFromGenerator}
\end{align}
where the ``generator'' is
\begin{align}
 {\sf G} &= \left(
  \begin{tabular}{cccc}
      0 & 0 & 0 & $-q^4/(8\pi)$ \\
      1 & 0 & $-q^2/(4\pi)$ & 0 \\
      0 & $8\pi$ & 0 & 0 \\
      0 & 0 & 1 & 0
  \end{tabular}\right).
  \label{EQgenerator}
\end{align}
As an almost trivial consequence of (\ref{EQkernelFromGenerator}), the
relationship
\begin{align}
  \tilde {\sf K}(R+R',q) =  \tilde {\sf K}(R,q) \cdot \tilde {\sf K}(R',q)
\end{align}
holds, simply due to the exponential satisfying $\exp((R+R'){\sf
  G})=\exp(R{\sf G})\exp(R'{\sf G})$. Correspondingly, in real space
\begin{align}
  {\sf K}(R+R',r) = {\sf K}(R) \;\dot\ast\; {\sf K}(R').
  \label{EQchainOfK}
\end{align}
Two features of this expression are welcome. One is that the the
Kierlik-Rosinberg weight functions $w_\nu(R,r)$ constitute four of the
sixteen ($4\times 4$) components of ${\sf K}(r,R)$.  Hence, the
Kierlik-Rosinberg weight functions are ``automatically'' generated
when starting with (\ref{EQkernelFromGenerator}) and
(\ref{EQgenerator}). The second feature is that all further components
$K_\nu^\tau(R,r)$ share their degree of non-locality with the weight
functions, i.e., all components of $\sf K$ vanish for distances
$r>R$. This becomes a crucial property when one uses these objects as
more general convolution kernels in the construction of third and
higher orders (in density) in the excess free energy
functional. Before we describe such developments we return to the
problem of a general binary hard sphere mixtures, in which the cross
interaction is non-additive, i.e., $\sigma_{12} \neq R_1+R_2 =
(\sigma_{11}+\sigma_{22})/2$.

Equation~(\ref{EQchainOfK}) allows us to use two convolutions rather than
only one in order to represent the cross Mayer function for a
non-additive binary mixture as
\begin{align}
  -f_{12}(r) = {\sf w}(R_1) \;\dot\ast\; {\sf K}(d) \;\dot\ast\; {\sf w}^t(R_2).
  \label{EQf12nonadditiveBinary}
\end{align}
Here, the length scale $d$ is due to the non-additivity, and satisfies
$\sigma_{12}=R_1+d+R_2$. Figure~\ref{FIGone}~(d) shows the
corresponding diagrams, where ${\sf K}(d)$ constitutes an additional
type of bond. The length of the bond can be adjusted, via changing
$d$, in order to generate the (given) range of $f_{12}(r)$.  In the
diagram, the position of each kink is integrated over. These integrals
correspond to the convolution integrals in
(\ref{EQf12nonadditiveBinary}). The low density expansion of the
binary non-additive hard sphere functional~\cite{schmidt04nahs}
contains the same interspecies diagrams as the additive
Kierlik-Rosinberg functional, as shown in figure~\ref{FIGone}~(a) and
(b), and contains the diagram shown in figure~\ref{FIGone}~(d) for the
cross species interaction.

Using the property (\ref{EQchainOfK}), one can go beyond
(\ref{EQf12nonadditiveBinary}) and represent the single $\sf K$ matrix
via further deconvolution as a (convolution) product of two $\sf K$
matrices, with appropriately chosen length scales $d_1$ and $d_2$,
so that $\sigma_{12}=R_1+d_1+d_2+R_2$, and hence
\begin{align}
  -f_{12}(r) = {\sf w}(R_1) \;\dot\ast\; {\sf K}(d_1) \;\dot\ast\;
   {\sf K}(d_2) \;\dot\ast\; {\sf w}^t(R_2).
  \label{EQf12nonadditiveTernary}
\end{align}
The corresponding diagram is shown in figure~\ref{FIGone}~(e). This
constitutes the cross species low-density limit of the ternary
non-additive hard sphere functional~\cite{schmidt11tnas}. For ternary
mixtures, the three cross species lengthscales $\sigma_{12},
\sigma_{13}, \sigma_{23}$ are decomposed as
\begin{align}
  \sigma_{ij} = R_i + d_i + d_j + R_j\,,
\end{align}
with appropriate values of $d_1,d_2$ and $d_3$, which can be uniquely
determined for a mixture with three components
\cite{schmidt11tnas}.

\begin{figure}[ht]
\centering
\includegraphics[width=\myfigwidth]{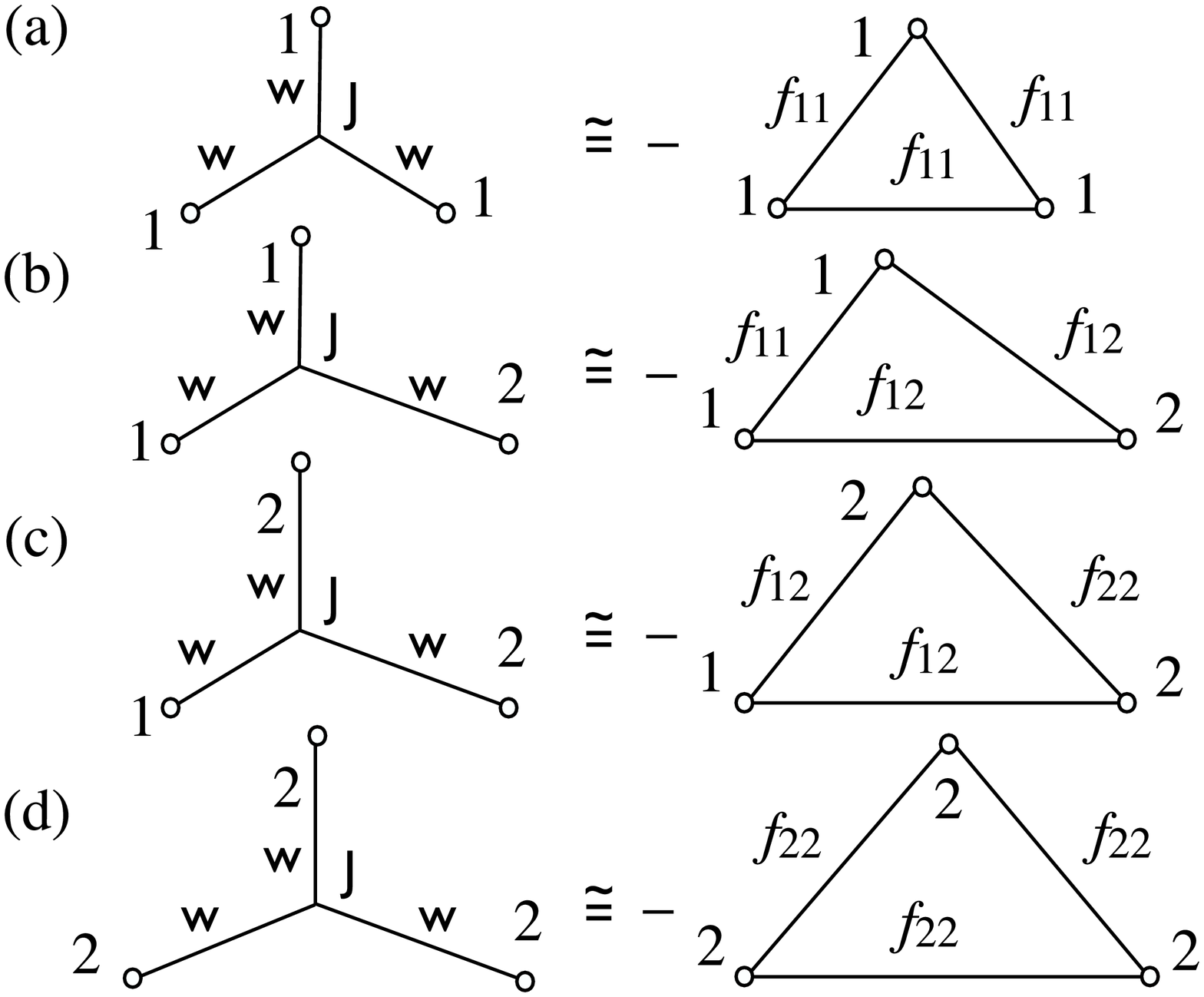}
\caption{Diagrams of the FMT for additive hard sphere mixtures at the
  third virial level. On the right hand side, the exact third order
  diagrams are shown for a mixture of two species, 1 and 2, where the
  bonds are Mayer functions $f_{ij}(r)$ between species $i,j=1,2$. The
  left hand sides show the corresponding FMT approximation. Weight
  function vectors $\sf w$ for species 1 and 2 are connected to a
  central junction. The position of the central junction is integrated
  over, and the geometric index of the weight functions is
  tensor-contracted with the third-rank ``junction tensor'' $\sf J$.}
\label{FIGtwo}
\end{figure}

We turn to the third-virial level. The exact contribution to the
excess free energy functional is
\begin{align}
  -\frac{k_{\mathrm{B}}T}{6}\sum_{ijk}
  \int \rd\rv \int \rd\rv' \int \rd\rv''
  \rho_i(\rv) \rho_j(\rv') \rho_k(\rv'')
  %
f_{ij}(\rv-\rv') f_{jk}(\rv'-\rv'') f_{ik}(\rv-\rv'').
  \label{EQfexcThirdVirial}
\end{align}
Due to the pairwise coupling of the three integration variables, one
commonly refers to (\ref{EQfexcThirdVirial}) as (the sum of) triangle
diagrams. FMT fails to generate this exact expression~\cite{cuesta02},
but yields a very reasonable approximation to it. Apart from going from
Mayer bonds as convolution kernels to weight function bonds, the
crucial step is a ``re-wiring'' or topological change~\cite{korden2012virial} in the structure of the diagrams. Avoiding the
loop in the triangle diagram (a problem that gets more severe with
an increasing order in density in the virial expansion, see below), the
FMT diagrams connect the (three) weight function bonds to a {\em
  common} central space integral. A tree-like (or star-like) topology
results, see figure~\ref{FIGtwo}. Writing out the space integrals
explicitly yields
\begin{align}
 - f_{ij}(\rv-\rv') f_{jk}(\rv'-\rv'') f_{ik}(\rv-\rv'')
  \approx  \sum_{\nu,\tau,\kappa=0}^3 J^{\nu\tau\kappa}
  %
  %
  \int \rd\xv
   w_\nu(R_i,\rv-\xv)
   w_\tau(R_j,\rv'-\xv)
   w_\kappa(R_k,\rv''-\xv).
   \label{EQtriangleApproximation}
\end{align}
From an algebraic point of view, the operation on the right hand side
of (\ref{EQtriangleApproximation}) is similar to a triple scalar
product of ${\sf w}(R_i)$, ${\sf w}(R_j)$, and ${\sf w}(R_k)$. The
third-rank tensor $\sf J$ that generates this operation can be derived~\cite{leithall11hys} from the sole requirement that the result, after
the $\xv$-integral in (\ref{EQtriangleApproximation}), does not
contain unphysical divergences (more precisely that the corresponding
order in the partial bulk direct correlation function is finite
everywhere). The tensor $\sf J$ is symmetric in all its indices; its
non-vanishing components are $J^{033}=J^{123}=1$, and
$J^{222}=1/(4\pi)$.

\begin{figure}[!b]
\centering
\includegraphics[width=\myfigwidth]{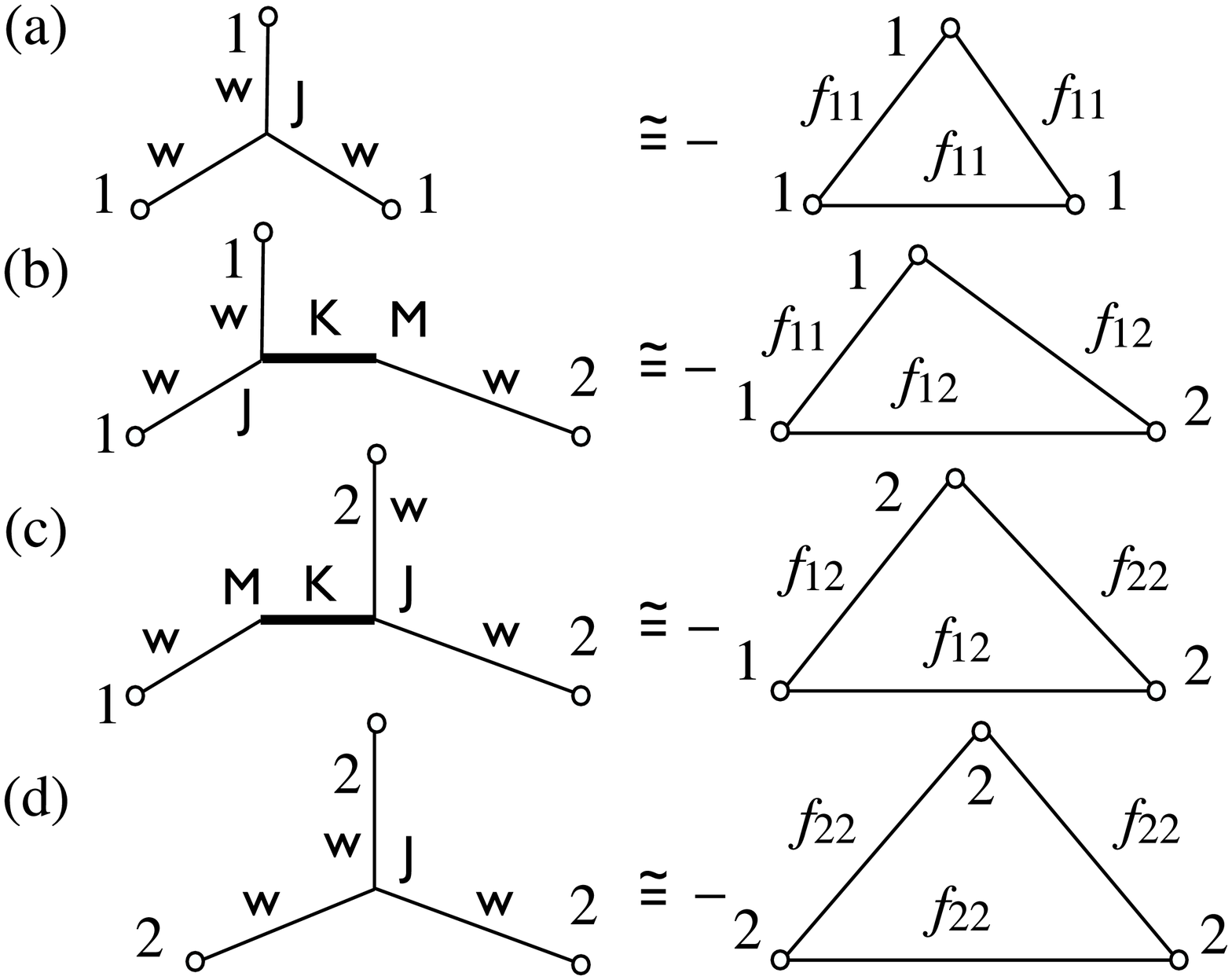}
\caption{Same as figure~\ref{FIGtwo}, but for binary non-additive hard
  sphere mixtures. The diagrams that involve either species 1 (a) or
  species 2 (d) exclusively are the same as those for the additive
  mixture, compare to figure~\ref{FIGtwo}~(a) and (d). The mixed FMT diagrams
  (b) and (c) feature the kernel matrix $\sf K$ as an additional type
  of bond. $\sf K$ is a second rank tensor, which is contracted via an
  $\sf M$ metric to a $\sf w$ bond and, on its other side to the
  third-rank junction tensor $\sf J$. This allows one to control the range
  of the cross Mayer bond $f_{12}(r)$ freely. The $\sf w$ bonds are specific for each species (indicated
  by the species index 1,2 at each root point).}
\label{FIGthree}
\end{figure}

Figure~\ref{FIGtwo} illustrates these diagrams for the case of an
additive binary mixture. Note that, as on the second virial level, all
bond lengths are pre-determined, once the particle radii $R_1$ and
$R_2$ have been chosen. These are uniquely determined from
$R_i=\sigma_{ii}/2$, leaving no room to go beyond the additive
case. See~\cite{cuesta02} for an in-depth discussion of the so-called
``lost cases'' of FMT that come from the intrinsic differences between
the triangle and the three-arm star diagrams. In short, when
stretching all bonds maximally, the exact diagram is larger than the
FMT approximation; see~\cite{cuesta02}. Nevertheless, the different
lengths of the FMT arms approximate the equilateral cases $ijk=111$,
figure~\ref{FIGtwo}~(a) and $ijk=222$, figure~\ref{FIGtwo}~(d), as well
as simultaneously the isosceles cases $ijk=112$, figure~\ref{FIGtwo}~(b) and $ijk=122$, figure~\ref{FIGtwo}~(c). For
non-additive mixtures, however, this is is insufficient for
approximating all triangle diagrams.

For non-additive mixtures, $\sf K$ can be used in order to replace one
of the weight functions ${\sf w}(R)$ in
(\ref{EQtriangleApproximation}) by ${\sf K}(d)\;\dot\ast\; {\sf
  w}(R)$.  The resulting diagrams are shown in figure~\ref{FIGthree}.
The intra-species contributions, $ijk=111$ in figure~\ref{FIGthree}~(a)
and $ijk=222$ in figure~\ref{FIGthree}~(d), are unchanged as compared
to the additive case shown in figure~\ref{FIGtwo}. The inter-species
diagrams differ by an additional $\sf K$ bond that acts as a spacer
between the arm of the ``minority''-species (i.e., the one that
appears only once, not twice) and the central space integral. Note
that both ``majority'' arms directly connect to the center. This
applies to $ijk=112$, as shown in figure~\ref{FIGthree}~(b), and to
$ijk=122$ as shown in figure~\ref{FIGthree}~(c).

\begin{figure}[ht]
\centering
\includegraphics[width=\myfigwidth]{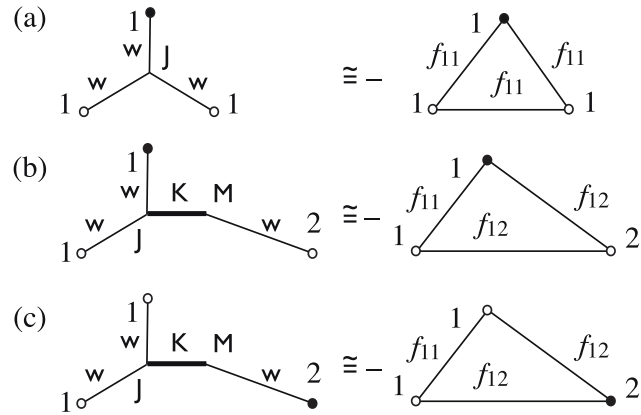}
\caption{When one field point is integrated over (with no or a
  constant density) the approximate FMT diagrams give the exact
  result. Here, a selection of the relevant diagrams is shown. The
  open symbols are root points, i.e., fixed position arguments. The
  filled symbols are field points that are integrated
  over. Representative examples are shown: the diagram in (a) is for
  species $ijk=111$, (b) and (c) are for species 112, where in (b) a
  1-field point and in (c) a 2-field point is integrated over.}
\label{FIGfour}
\end{figure}

Although the FMT expressions for third-order diagrams are
approximations, they possess one important property which is most
clearly analysed when considering the low-density limit of the
(partial) bulk fluid two-body direct correlation functions,
$c_{ij}(r)$, obtained as second functional derivatives,
\begin{align}
  c_{ij}(|\rv-\rv'|) = -\frac{1}{k_{\mathrm{B}}T} \left.\frac{\delta^2 F_{\rm
      exc}[\{\rho_k\}]}{\delta\rho_i(\rv)\delta\rho_j(\rv')}
  \right|_{\rho_l={\rm const}}.
  \label{EQcijFromDerivative}
\end{align}
Observing (\ref{EQfexc2ndOrder}) and (\ref{EQfexcThirdVirial})
\begin{align}
  c_{ij}(r) = f_{ij}(r) + \sum_k \rho_k c^*_{ijk}(r) + O(\rho_l^2),
\end{align}
where the contribution linear in densities is obtained from the
triangle diagram by multiplying one of the root points with the bulk
density and integrating its position over space, i.e., turning the
root point of species $k$ into a density field point. The field is
constant in the homogeneous bulk.  A selection of the corresponding
integrals is displayed in figure~\ref{FIGfour}. The FMT results give
the exact result for $c_{ijk}^*(r)$ for all combinations of
species. This applies to all versions of the theory, whether additive
\cite{rosenfeld89,kierlik90}, binary non-additive
\cite{schmidt04nahs}, or ternary non-additive~\cite{schmidt11tnas},
and is a first indication of successful description of bulk structure.

\begin{figure}[ht]
\centering
\includegraphics[width=\myfigwidth]{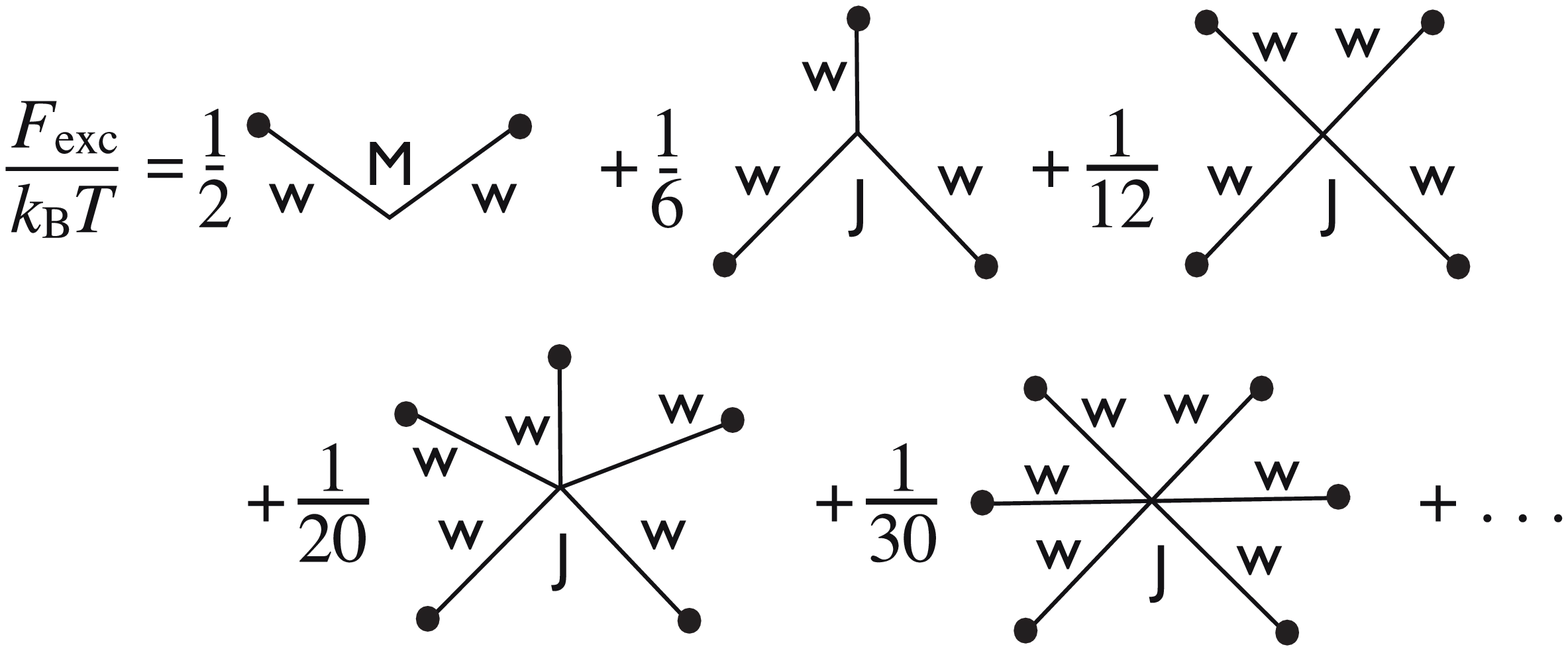}
\caption{Diagrammatic series of the Kierlik-Rosinberg form of the
  additive fundamental measure free energy functional. The field
  points (filled circles) represent one-body densities and are
  integrated over. The bonds are fundamental measure weight functions
  that are joined by an $\sf M$ metric (second order in density), or
  by junction tensor $\sf J$ of $k$-th rank ($k$-th order in
  density). The position of the inner junction (center of the star) is
  integrated over. This corresponds to the ``outer'' integral $\int
  \rd{\bf x}$ over the free energy density $\Phi(n_\tau({\bf x}))$ in
  the standard, weighted density formulation with weighted densities
  $n_\tau({\bf x})$. The scalar coefficient of the $k$-th order
  diagram is $1/[k(k-1)]$.}
\label{FIGfive}
\end{figure}

The construction of the additive hard sphere FMT free energy
functional can be based on a diagrammatic series using the assumption
that all diagrams of higher than third order in density are also of
star-like shape~\cite{leithall11hys}. The weight functions $\sf w$
that constitute the arms are connected to a single central space
integral.  In order to connect $p$ arms of the $p$-th order in density,
a tensor of rank $p$ is required.  One can show~\cite{leithall11hys}
that a recursive relation holds and that these object can be obtained
by tensor contraction of an appropriate number ($p-2$) of third-rank
tensors $\sf J$. As an illustration, at fourth order
$J^{\nu\nu'\tau\tau'} = \sum_{\kappa=0}^3 J^{\nu\nu'}_\kappa
J^{\kappa\tau\tau'}$, where the first tensor on the right hand side
has one index lowered via contraction with the metric,
$J^{\nu\nu'}_\kappa = \sum_{\kappa'=0}^3 M_{\kappa\kappa'}
J^{\nu\nu'\kappa'}$.  Together with the coefficients $1/[p(p-1)]$,
which are taken from the zero-dimensional properties of the system
\cite{leithall11hys}, the resulting series is displayed in figure~\ref{FIGfive}. The series can be explicitly summed, and yields, in
three spatial dimensions, the Kierlik-Rosinberg form of FMT. Each arm
together with its filled field point constitutes a ``weighted
density''~\cite{rosenfeld89,kierlik90},
\begin{align}
  n_\tau(\xv) = \int \rd\rv \rho(\rv)w_\tau(\rv-\xv).
\end{align}

\begin{figure}[!h]
\centering
\includegraphics[width=\myfigwidth]{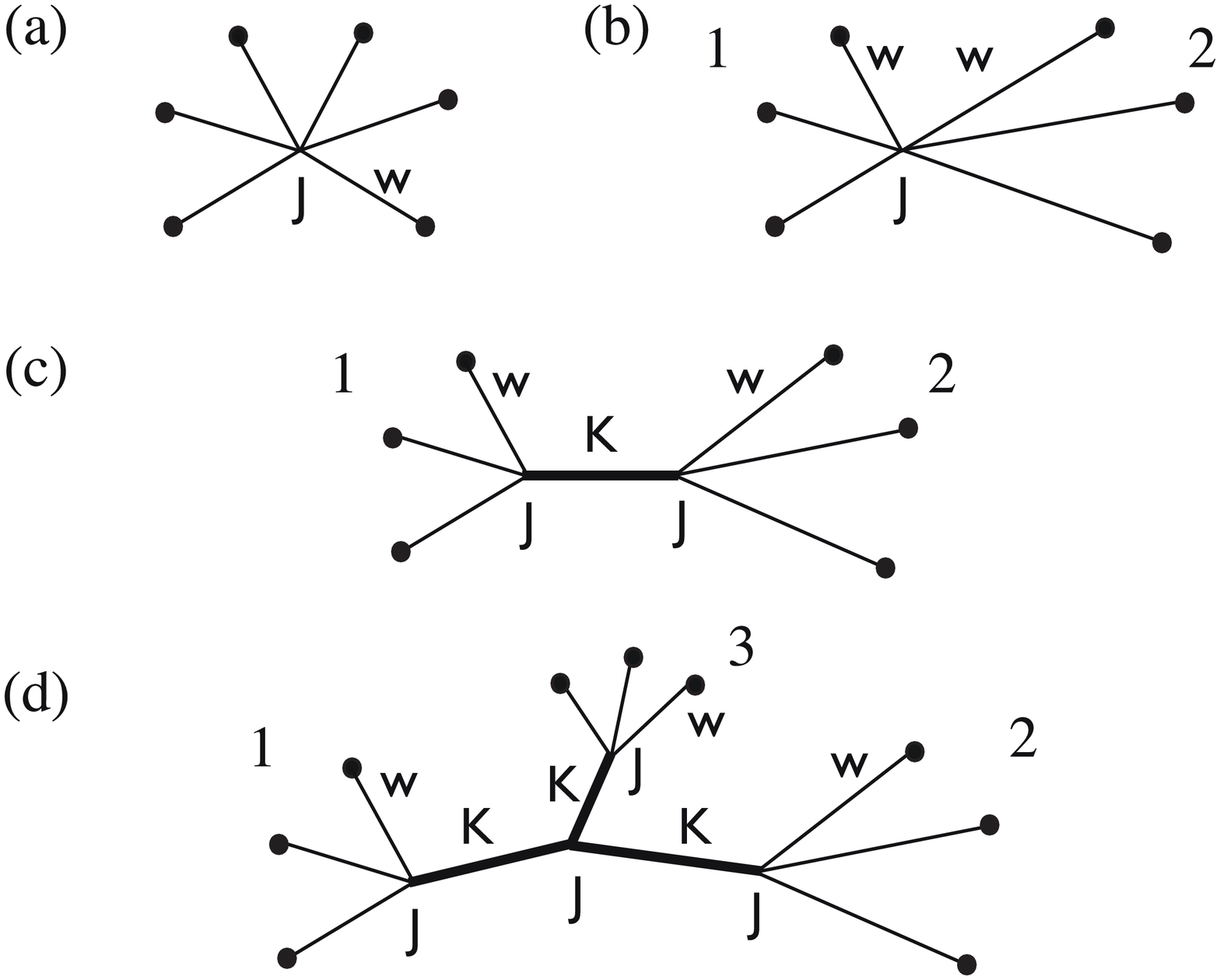}
\caption{Examples of the topology of the higher (than third) order
  diagrams that constitute the respective FMT free energy
  functional. Shown are topologies for the pure system (a), additive
  binary mixtures (b), non-additive binary mixtures (c), and
  non-additive ternary mixtures (d). Contributions of sixth order in
  density are shown in (a)--(c), and of ninth order in density in (d).}
\label{FIGsix}
\end{figure}

The diagrams that constitute the FMTs for binary and
ternary non-additive hard sphere mixture, possess different topology; a
summary is displayed in figure~\ref{FIGsix}. The one-component hard
sphere FMT functional, shown in figure~\ref{FIGsix}~(a) consists of
stars with arms of the same length $R$. In one dimension (hard rods
on a line), this is equivalent to Percus' exact functional
\cite{percus76}. The binary version possesses two different types of
arms corresponding to the small (here species 1) and large (species 2)
component, see figure~\ref{FIGsix}~(b).  The one-dimensional version of
the functional for non-additive hard core mixtures
\cite{schmidt07naho} is an approximation that differs from the exact
solution~\cite{santos07oned}.  The binary non-additive FMT has two
central junctions, where the arms of each of them belong to one
species only (the small species 1 belongs to the left center, and the
large species 2 to the right center). The two centers are connected
via a $\sf K$ bond of appropriate length $d$, so that
$\sigma_{12}=R_1+d+R_2$, see figure~\ref{FIGsix}~(c). Finally, figure~\ref{FIGsix}~(d) displays a typical diagram of the ternary non-additive
FMT. Here, there are three subcenters, one for each species, with
corresponding arms. The subcenters are connected via $\sf K$ bonds to
one global center.  The positions of all junctions are integrated
over.  A full account of the ternary FMT functional can be found in
~\cite{schmidt11tnas}.

\subsection{Non-additive hard sphere fluids in bulk and at interfaces}
\label{SECapplications}

We give a brief summary of applications of the FMT for binary
non-additive hard sphere mixtures. Based on the FMT for this system
\cite{schmidt04nahs,schmidt11tnas} but also on Monte Carlo computer
simulations, a range of physical phenomena were investigated. The bulk
fluid structure on the two-body level is described very satisfactorily
by the theory, as compared to simulation data. This applies to the results
for the partial pair correlation functions, $g_{ij}(r)$, when obtained
through the Ornstein-Zernike route using the partial direct
correlations functions, $c_{ij}(r)$, obtained as the second functional
derivative (\ref{EQcijFromDerivative}) of the free energy
functional. This is a quite severe test of the theory, because  both the
derivative and the Ornstein-Zernike equations constitute involved
operations. The general performance of the theory is very
satisfactory, although the core condition $g_{ij}(r<\sigma_{ij})=0$ is
only approximately satisfied. This can be rectified with Percus'
test-particle method~\cite{percus62}, where the grand potential is
minimized in the presence of an external potential that is equal to
the inter-particle interaction. Test-particle results for the
$g_{ij}(r)$ reproduce the simulation data very well
\cite{hopkins11natpl}, as long as the system is away from the
``decoupling case'' of negative non-additivity, so that
$\sigma_{12}\ll(\sigma_{11}+\sigma_{22})/2$. It is important to note
that the theory does not yield unphysical artifacts in test-particle
results for partial pair correlation functions
\cite{ayadim10,hopkins11natpl}

The theory yields analytic expressions for the partial direct
correlation functions (both in real space and in Fourier space), and
hence, via the Ornstein-Zernike relation also analytic expressions for
the partial structure factors $S_{ij}(q)$. This makes it very
convenient to carry out an analysis of the asymptotic, large distance
decay of bulk pair correlation functions from pole analysis of the
$S_{ij}(q)$ in the plane of complex wavevectors $q$
\cite{evans93,evans94}, and to relate these results to the decay of
one-body density profiles in inhomogeneous situations, such as at
interfaces~\cite{hopkins10nahs}.

For positive non-additivity,
$\sigma_{12}>(\sigma_{11}+\sigma_{22})/2$, which is sufficiently
large, the system displays fluid-fluid phase separation into two fluid
phases with different chemical compositions. The theory gives good
account of the location of the fluid-fluid demixing binodal in the
plane of partial packing fractions of the two species
\cite{schmidt04nahs,hopkins10nahs}. A wealth of interesting interfacial
phenomena results as a consequence of the bulk fluid demixing, and we
refer the reader to the original papers on fluid demixing, asymptotic
decay of correlations and free fluid interfaces~\cite{hopkins10nahs},
first-order layering and critical wetting transitions in non-additive
hard sphere mixtures~\cite{hopkins11nawe}, and capillary condensation
of non additive hard sphere mixtures in planar confinement
\cite{hopkins12naslt}.

\section{Conclusions}
\label{SECconclusions}

We have described a range of recent developments and applications of
classical density functional theory in the study of bulk and
interfacial properties of liquids. Starting with a reassessment of the
underlying variational principle, we have laid out how to use Levy's
constrained search method in order to define the free-energy
functional.  Levy's method provides us with an explicit expression for
the free-energy functional (\ref{EQFviaLevy}). We showed that the
concept can be generalized in order to define an internal-energy
functional, which possesses the one-body density distribution and a
local entropy distribution as trial fields. A dynamical theory built
on the internal-energy functional can be found in
\cite{schmidt11edft}. Using Levy's method, the definition of the
functional is explicitly independent of the external potential, which
constitutes a conceptual advantage. However, the minimization in the
function space of many-body probability distributions cannot be in practice
 carried out for a realistic model Hamiltonian. Hence, one has to
rely on approximations for the functional, as is common in DFT.  We
refer the reader to~\cite{schmidt11edft} for a description of various
approximate internal-energy functionals, which have been obtained from
Legendre transforming the corresponding approximation for the
Helmholtz free-energy functional.

We have laid out the basic ideas underlying the recent progress in
formalizing the mathematical structure of FMT. This includes the
tensorial nature, in that the ``geometric'' indices of the
Kierlik-Rosinberg weight functions play the role of tensorial
indices. A diagrammatic notation helps to clarify the non-local nature
of the excess free energy functional. We have given an overview of
applications of the binary non-additive hard sphere functional to a
range of phenomena.

The above work extends the previous efforts that were primarily based on
the intimate connection of the properties of the free energy
functional under dimensional crossover, i.e., the result of applying
the functional to density distributions that correspond to an extreme
confinement in one or more spatial directions, made it possible to generalize
FMT to the Asakura-Oosawa-Vrij model~\cite{asakura54,vrij76} of
colloid-polymer mixtures~\cite{schmidt00cip}, the Widom-Rowlinson
model~\cite{schmidt01wr}, and penetrable spheres that interact with a
constant repulsive plateau~\cite{schmidt99ps}. These models have in
common that their zero-dimensional properties, i.e., the statistical
mechanics of a cavity of the size of a single particle, carries still
some of the essentials of the true three-dimensional problem. Hence,
the zero-dimensional problem, which can be solved exactly (or with
good approximations~\cite{schmidt99ps}) in the above cases, is
sufficient as a central modification over the FMT for hard spheres, in
order to obtain a reliable free energy functional. See
e.g.,~\cite{brader02swet,schmidt03capc,schmidt04cape,fortini05tensnew,fortini06}
for applications to confined model colloid-polymer mixtures.

In future work it would be interesting to apply the dynamical test
particle limit~\cite{archer07dtpl,hopkins10dtpl} to non-additive hard
sphere mixtures in order to gain a better understanding of the
dynamical behaviour of such mixtures.  A further important problem is
to re-consider formulating an FMT for soft sphere
models~\cite{schmidt99sfmf,schmidt00fdec,schmidt00mix,groh00sfmt} in
the light of the diagrammatic and tensorial structure. Work along
these lines is in progress~\cite{burgis12}.  It would also be
interesting to investigate the implications of Levy's method for the
statistical mechanics of quenched-annealed fluid mixtures, where DFT
was obtained both via the replica trick
\cite{schmidt02pordf,schmidt02aom,reich04poroned,schmidt05liquidMatter},
and via a first-principles derivation following the Mermin-Evans
arguments~\cite{lafuente06qadft}.\\

\section*{Acknowledgement}
We thank R.~Evans, A.J.~Archer, J.-P.~Hansen, J.M.~Brader
for useful discussions. WSBD acknowledges an ORS award of the
University of Bristol. This work was supported by the EPSRC under
Grant EP/E065619/1 and by the DFG via SFB840/A3.


\ukrainianpart

\title{Нові досягнення в класичній теорії функціоналу густини:
       функціонал внутрішньої енергії і діаграмна структура теорії фундаментальної міри}
\author{М.~Шмідт\refaddr{bt,bs}, М. Бургіс\refaddr{bt},
 В.С.Б.~Двандару\refaddr{bs,indonesia}, Г. Ляйталь\refaddr{bs}, П. Гопкінс\refaddr{bs}}
\addresses{
\addr{bt} Відділення теоретичної фізики II, Інститут фізики, Університет Байройту, 95440 Байройт, Німеччина
\addr{bs} Фізична лабораторія Г.Г. Вілліс, Університет Бристолю, Бристоль BS8 1TL, Велика Британія
\addr{indonesia} Кафедра фізики, Державний університет  Джок'якарти, Джок'якарта, Індонезія
}

\makeukrtitle

\begin{abstract}
\tolerance=3000%
Зроблено огляд декількох недавніх праць з теорії функціоналу густини для класичних неоднорідних рідин.
Ми показуємо яким чином метод Леві обмеженого пошуку  може бути використаний для виведення варіаційного принципу, який
лежить в основі теорії функціоналу густини. Перевагою цього методу є те, що вільна енергія Гельмгольтца як функціонал
пробної одночастинкової густини задається у явному вигляді без відносно до зовнішнього потенціалу,  як це є у вападку
стандартного доведення Мерміна-Еванса через  {\it reductio ad absurdum}. Ми показуємо як узагальнити підхід  для того, щоб
виразити внутрішню енергію у вигляді функціоналів розподілів одночастинкової густини і локальної ентропії. Тут локальний хімічний потенціал і
температура в об'ємі  відіграють роль  множників  Лагранжа  в рівнянні Ейлера-Лагранжа для мінімізації функціоналу.
Як наближення для функціоналу вільної енергії,  показано діаграмну структуру  функціоналу
густини фундаментальної міри Розенфельда  для суміші твердих сфер. Описано недавні узагальнення для бінарних і потрійних сумішей твердих сфер, які
грунтуються на скалярних зважувальних функціях Кієрліка-Розінберга.

 \keywords теорія функціоналу густини, теорема Гогенберга-Кона, функціонал Розенфельда
\end{abstract}

\end{document}